\documentclass[12pt]{article}

\usepackage{amssymb}
\usepackage{amsmath}
\usepackage{amsfonts}
\usepackage{graphicx}
\usepackage{mathrsfs}
\usepackage{comment}
\usepackage{cite}

\begin{document}

\renewcommand{\figurename}{Fig.}
\renewcommand{\tablename}{Table.}
\newcommand{\Slash}[1]{{\ooalign{\hfil#1\hfil\crcr\raise.167ex\hbox{/}}}}
\newcommand{\bra}[1]{ \langle {#1} | }
\newcommand{\ket}[1]{ | {#1} \rangle }
\newcommand{\beq}{\begin{equation}}  \newcommand{\eeq}{\end{equation}}
\newcommand{\bef}{\begin{figure}}  \newcommand{\eef}{\end{figure}}
\newcommand{\bec}{\begin{center}}  \newcommand{\eec}{\end{center}}
\newcommand{\non}{\nonumber}  \newcommand{\eqn}[1]{\begin{equation} {#1}\end{equation}}
\newcommand{\laq}[1]{\label{eq:#1}}  
\newcommand{\dd}[1]{{d \o d{#1}}}
\newcommand{\Eq}[1]{Eq.(\ref{eq:#1})}
\newcommand{\Eqs}[1]{Eqs.(\ref{eq:#1})}
\newcommand{\eq}[1]{(\ref{eq:#1})}
\newcommand{\Sec}[1]{Sec.\ref{chap:#1}}
\newcommand{\ab}[1]{\left|{#1}\right|}
\newcommand{\vev}[1]{ \left\langle {#1} \right\rangle }
\newcommand{\bs}[1]{ {\boldsymbol {#1}} }
\newcommand{\lac}[1]{\label{chap:#1}}
\newcommand{\SU}[1]{{\rm SU{#1} } }
\newcommand{\SO}[1]{{\rm SO{#1}} }
\def\({\left(}
\def\){\right)}
\def\dt{{d \o dt}}
\def\diag{\mathop{\rm diag}\nolimits}
\def\Spin{\mathop{\rm Spin}}
\def\O{\mathcal{O}}
\def\U{\mathop{\rm U}}
\def\Sp{\mathop{\rm Sp}}
\def\SL{\mathop{\rm SL}}
\def\tr{\mathop{\rm tr}}
\def\ebq{\end{equation} \begin{equation}}
\newcommand{\OR}{~{\rm or}~}
\newcommand{\AND}{~{\rm and}~}
\newcommand{\EV}{ {\rm ~eV} }
\newcommand{\KEV}{ {\rm ~keV} }
\newcommand{\MEV}{ {\rm ~MeV} }
\newcommand{\GEV}{ {\rm ~GeV} }
\newcommand{\TEV}{ {\rm ~TeV} }
\def\o{\over}
\def\a{\alpha}
\def\b{\beta}
\def\c{\varepsilon}
\def\d{\delta}
\def\e{\epsilon}
\def\f{\phi}
\def\g{\gamma}
\def\h{\theta}
\def\k{\kappa}
\def\l{\lambda}
\def\m{\mu}
\def\n{\nu}
\def\p{\psi}
\def\q{\partial}
\def\r{\rho}
\def\s{\sigma}
\def\t{\tau}
\def\u{\upsilon}
\def\v{\varphi}
\def\w{\omega}
\def\x{\xi}
\def\y{\eta}
\def\z{\zeta}
\def\D{\Delta}
\def\G{\Gamma}
\def\H{\Theta}
\def\L{\Lambda}
\def\F{\Phi}
\def\P{\Psi}
\def\S{\Sigma}
\def\me{\mathrm e}
\def\ol{\overline}
\def\tl{\tilde}
\def\*{\dagger}

\begin{center}

\hfill MIT-CTP/5022\\
\hfill   TU-1064\\
\hfill  IPMU18-0088\\

\vspace{1.5cm}

{\Large\bf The QCD Axion Window and Low-Scale Inflation }
\vspace{1.5cm}

{\bf Fuminobu Takahashi$^{1,2,3}$, Wen Yin$^{4}$, Alan H. Guth$^{3}$}

\vspace{12pt}
\vspace{1.5cm}
{\em 
$^{1}$Department of Physics, Tohoku University,  
Sendai, Miyagi 980-8578, Japan \\
$^{2}$Kavli Institute for the Physics and Mathematics of the Universe (WPI),
University of Tokyo, Kashiwa 277--8583, Japan\\
$^{3}$Center for Theoretical Physics \& Department of Physics, Massachusetts Institute of Technology, Cambridge, MA 02139, USA\\
$^{4}$ Institute of High Energy Physics, Chinese Academy of Sciences, Beijing 100049,  China \vspace{5pt}}

\vspace{1.5cm}
\abstract{
We show that the upper bound of the classical QCD axion window can be significantly relaxed 
for low-scale inflation. If the Gibbons-Hawking temperature during inflation is lower
than the QCD scale, the initial QCD axion misalignment angle follows the Bunch-Davies distribution.
The distribution is peaked at the strong CP conserving minimum if there is no other light degree of freedom 
contributing to the strong CP phase. 
As a result, the axion overproduction problem is significantly relaxed even for an axion decay constant 
larger than $10^{12}\GEV$. We also provide concrete hilltop inflation models where the Hubble parameter
during inflation is comparable to or much smaller than the QCD scale, with successful reheating taking place
via perturbative decays or dissipation processes. 
}

\end{center}
\setcounter{footnote}{0}
\section{Introduction}

The QCD axion is a pseudo Nambu-Goldstone boson associated with the spontaneous breakdown of
a global Peccei-Quinn symmetry \cite{Peccei:1977hh, Peccei:1977ur, Weinberg:1977ma}. Interestingly, the QCD axion not only solves the strong CP problem,
but also explains dark matter as it is copiously produced by the initial 
misalignment mechanism in the early universe~\cite{Preskill:1982cy,Abbott:1982af,Dine:1982ah}.

The axion is massless at high temperatures, but it acquires a mass from 
non-perturbative effects of QCD at low energy, solving the strong CP problem. 
In the early universe, therefore, there is no  reason for the QCD axion to sit
exactly at the CP conserving minimum, and it is usually assumed that the initial position, $a_*/f_a$, 
is of order of unity, where $f_a$ is the axion decay constant.
Then, the QCD axion starts to oscillate about the CP conserving minimum 
around the QCD phase transition, and the coherently oscillating axion becomes dark matter.
See e.g. \cite{Kim:2008hd,Wantz:2009it,Ringwald:2012hr,Kawasaki:2013ae,Marsh:2015xka} for recent reviews. 

The so-called classical axion window is given by
\begin{align}
\label{aw}
4 \times 10^8{\rm\,GeV} \lesssim f_a \lesssim 10^{12}\,{\rm GeV},
\end{align}
where the upper bound is due to the axion abundance described above~\cite{Preskill:1982cy,Abbott:1982af,Dine:1982ah},
and the lower is due to the neutrino burst duration of SN1987A~\cite{Mayle:1987as,Raffelt:1987yt,Turner:1987by}.
Therefore, if the axion decay constant is of order the GUT scale or string scale, $f_a\sim 10^{16-17}$\,GeV,
the axion abundance exceeds the observed dark matter abundance unless the initial position $a_*/f_a$ is fine-tuned
to be much smaller than unity. 

There are several ways to relax the upper bound of the classical axion window. For instance, the axion
abundance can be diluted by the late-time entropy production after the QCD phase 
transition~\cite{Dine:1982ah,Steinhardt:1983ia,Lazarides:1990xp,Kawasaki:1995vt,Kawasaki:2004rx}. 
A small value of the initial misalignment angle may be selected in the multiverse based on the anthropic 
principle~\cite{Linde:1991km,Wilczek:2004cr,Tegmark:2005dy}.
Alternatively, the axion can acquire a time-dependent mass through
the Witten effect, if it is coupled to hidden monopoles~\cite{Kawasaki:2015lpf,Kawasaki:2017xwt}. If the effect is sizable,
the axion follows the time-dependent minimum adiabatically, and no particle production takes place at the time of the
QCD phase transition, suppressing the final abundance. Also, if there is a resonant mixing with 
axion-like particles, the axion abundance can be suppressed by the mass ratio~\cite{Kitajima:2014xla,Daido:2015bva,Daido:2015cba}.
More recently, it was pointed out that, if the axion has a large coupling to massless hidden photons, the axion energy
density is dissipated to hidden photons through tachyonic resonance, and the abundance can be suppressed 
by ${\cal O}(100)$~\cite{Kitajima:2017peg} (see also Ref.~\cite{Agrawal:2017eqm}).

In this Letter we propose another simple way to relax the upper bound of the axion window.
We show that, if the Hubble parameter during inflation is comparable to or lower than the QCD scale, 
the axion already acquires a nonzero mass during inflation,
and the distribution of its initial position follows the Bunch-Davies distribution~\cite{Bunch:1978yq}.
The distribution is peaked at the CP conserving minimum if there is no other light degree of freedom contributing to the strong CP phase. 
In other words, the typical initial misalignment angle is actually given by a function of the Hubble parameter during inflation.
As a result, the upper bound of the axion window can be significantly relaxed for such a low-scale inflation. 
We also provide concrete low-scale inflation models
 where successful reheating takes place via perturbative decays or dissipation processes.

\section{Bunch-Davies Distribution}
\label{BD}

In this section we briefly review the Bunch-Davis distribution of a light scalar field 
$\phi$ during de Sitter universe or eternal inflation with the Hubble parameter $H_{\rm inf}$. 
The action is given by 
\begin{equation}
S=\int{d^4x \sqrt{-g} \( - {1\over 2} g^{\mu \nu} {\partial \phi\over \partial x^\mu} {\partial \phi\over \partial x^\nu}-V(\phi)\)}.
\end{equation}
For simplicity we assume that the potential for the scalar is given by 
\begin{equation}
V(\phi)\simeq {m_\phi^2\over 2 }\phi^2 +V_0,
\end{equation}
where $V_0$ is the positive energy density leading to an exponential expansion of the universe,
and $m_\phi (>0) $ is the mass of the scalar. 
Note that $\phi$ is not the inflaton. 
Here we neglect the time dependence of $V_0$, which would decrease slowly in the case of the usual slow-roll inflation.  
The Hubble parameter is given by
\begin{equation}
H_{\rm inf}\equiv \sqrt{V_0 \over 3 M^2_{\rm pl}},
\end{equation}
where $M_{\rm pl}\simeq 2.4\times 10^{18}\GEV$ is the reduced Planck mass.
We assume that the scalar mass is lighter than the Hubble parameter,
\begin{equation} 
 m_\phi\ll H_{\rm inf}.
\end{equation}

We express the scalar field as $\phi({\bf x},t) =  \delta\phi({\bf x},t)+ \phi_0(t)$, and expand the space-dependent part
in the Fourier expansion as
\begin{equation}
\delta\phi({\bf x},t)= \int{{d^3{\bf k}\over (2\pi)^{3\over 2}} \left[ \delta\phi_k(t) \alpha_{\bf k} e^{i {\bf k \cdot x}} +\delta\phi^*_k(t) \alpha_{\bf k}^\dagger  e^{-i {\bf k \cdot x}} \right]},
\end{equation}
where $k=|\bf{k}|$, and $\alpha_{\bf k}$  and $\alpha^\dagger_{\bf k}$ will be later identified with the annihilation and creation operators, respectively. 
By solving the classical equation of motion for $\phi_0$, one can see that $\phi_0$ exponentially approaches zero as inflation ends
\begin{equation}
\phi_{0}(t_f)\simeq e^{-N{ m_\phi^2 \over 3H_{\rm inf}^2 }} \phi_{0}(t_{i}),
\end{equation}
where $N$ is the e-folding number for inflation, and $t_i$ and $t_f$ are the initial and final cosmic time during inflation. 
Assuming $N m_\phi^2/3H_{\rm inf}^2 \gg 1$, 
 $\phi_{0}(t_f)$  asymptotes to zero.
 On the other hand, as we will see below, the quantum fluctuation $\delta \phi$ will be accumulated during inflation,
 which will dominate over $\phi_{0}$. Therefore we will simply drop the zero mode by setting $\phi_0 = 0$ in the following analysis.

The equation of motion for $\delta\phi_k$ reads 
\begin{equation}
\delta\ddot{\phi_{ k}}+3H \delta\dot{\phi}_{ k}+ \left[m_\phi^2 +\({{ k}\over a}\)^2\right]{\delta \phi}_{ k} =0,
\end{equation}
where $a(t)=a_0\, e^{H_{\rm inf}t}$ is the scale factor.  
It is convenient to use the conformal time defined by
\begin{equation}
\eta\equiv \int^t{dt\over a(t)}=-{1\over aH}+ C
\end{equation}
where $C$ is a constant. We take hereafter $C=0$ so that 
\begin{align} 
\eta\rightarrow \left\{\begin{array}{ll}
	\displaystyle{-\infty ~~~~~{\rm as~~}  t\rightarrow -\infty}\\
	\displaystyle{~~ 0 ~~~~~~~{\rm as~~}t\rightarrow \infty}
	\end{array} \right..
\end{align}
The equation of motion can be rewritten as
\begin{equation}
{1 \over  k^2} {d^2 X_{ k} \over d \eta^2} +{1\over { k} \eta}{1 \over { k}} {d X_{ k} \over d \eta} +  
\left[1- \( {9\o4}-{m_\phi^2\over H_{\rm inf}^2}\){1\over \({ k} \eta\)^2}  \right] X_{ k}=0,
\end{equation}
where we have defined $X_{ k} \equiv  (-\eta)^{-3/2} \delta\phi_{ k}$.
Using the Hankel function, one can express  the solution as
\begin{equation}
X_{ k}\propto H_\nu^{(1)}(-{ k} \eta)
\end{equation}
with $\nu \equiv \sqrt{\frac{9}{4} - \frac{m_\phi^2}{H_{\rm inf}^2}}$.
Then the mode function is given by
\begin{equation}
\delta\phi_{ k}= {\sqrt{\pi} \over 2} H_{\rm inf}(-\eta)^{3/2} H_{\nu}^{(1)}(-k \eta).
\end{equation}
Here we have adopted the normalization so that it matches with the flat-space time result in the subhorizon limit, $k\eta\rightarrow -\infty$.
Since the scalar mass is much smaller than the Hubble parameter, one arrives at
\begin{equation}
\ab{\delta\phi_{ k}(t)}^2 \simeq {H_{\rm inf}^2\over 2k^3}\( {k \over a(t) H_{\rm inf}}\)^{2 m_\phi^2\over 3H_{\rm inf}^2}
\end{equation}
on superhorizon scales, keeping only the leading order of $m_\phi^2/H_{\rm inf}^2$.

At the subhorizon scales, the scalar field can be quantized in a usual way. 
Well inside the Hubble horizon, one can neglect the effect of the gravity and quantize the scalar field as in the flat spacetime
based on the canonical quantization conditions. They are equivalent to imposing the following conditions on
 the annihilation/creation operators,
\begin{align}
&[\alpha_{\bf k}, \alpha_{\bf k'}]=0, \\
&[\alpha_{\bf k}, \alpha^\dagger_{\bf k'}]=\delta^{(3)}({\bf k}- {\bf k}').
\end{align}
Then, one can define the Bunch-Davies vacuum by $\alpha_{\bf k} \ket{0}=0$ for any ${\bf k}$ with $\vev{0|0}=1$\cite{Bunch:1978yq}.
At the end of inflation ($t=t_f)$, the fluctuations of the scalar
field on scales of order the horizon obey a Gaussian
distribution, with a variance $\vev{\phi^2(t_f)}$ given by
\begin{align}
\vev{\phi^2(t_f)}&\simeq \int_{a(t_i) H_{\rm inf} }^{a(t_f) H_{\rm inf} }{{d^3{\bf k}\over (2\pi)^3} \ab{\delta\phi_{ k}(t_f)}^2} \non \\
&={3H_{\rm inf}^4\over 8\pi^2 m_\phi^2} \left[1-\( {a(t_i) \o a(t_f)}
\)^{2m_\phi^2 \o 3 H^2_{\rm inf}} \right] \simeq \(\sqrt{3 \over 8 \pi^2} {H_{\rm inf}^2 \over m_\phi}\)^{2}.
\label{variance}
\end{align}
The integration is over the modes that exited the horizon during
the period of inflation, from time $t_i$ (when $k/a(t_i)=H_{\rm
inf}$) to $t_f$ (when $k/a(t_f)=H_{\rm inf}$).  The final
approximation is justified by the assumption that
\begin{equation}
\( {a(t_i) \o a(t_f)} \)^{2m_\phi^2 \o 3 H^2_{\rm inf}} = \exp \( - {2
m_\phi^2 \o 3 H_{\rm inf}^2 } N \) \ll 1 \ ,
\end{equation}
an assumption that we also made in discussing the behavior of
$\phi_0(t)$.  This approximation is equivalent to setting the
lower limit of integration in Eq.~(\ref{variance}) to zero.

If we focus on the dynamics of the scalar field after the horizon exit of the CMB scales,
the typical initial value of $\phi$ is given by
\begin{equation}
\phi_{ i} \approx \phi_{\rm rms} = \sqrt{3 \over 8 \pi^2} {H_{\rm inf}^2 \over m_\phi}.
\end{equation}
If we allow the initial field value to be fine-tuned, with a probability of $10\%$, $1\%$, or $0.1\%$,
then $\phi_i$ would be bounded by $|\phi_i| < \epsilon \phi_{\rm rms}$, with
\begin{align}
\epsilon \simeq \left\{\begin{array}{ll}
	\displaystyle{ 0.126 ~~~~  (10\%~ {\rm tuning})}\\
	\displaystyle{0.0125  ~~~ (1\%~{\rm tuning})}\\
	\displaystyle{0.00125  ~~ (0.1\%~{\rm tuning})}\\
	\end{array} \right.. 
	\end{align}
In summary, if the inflation lasts sufficiently long, the distribution of the light scalar field
is given by a function of the Hubble parameter during inflation and its mass. Even if
the mass is lighter than the Hubble parameter, the scalar field knows where the potential
minimum is in a probabilistic way. This will be essential when we apply this result
to the QCD axion in the next section.

\section{The QCD axion abundance}
Now let us turn to the QCD axion abundance in a very low-scale inflation.
We identify the scalar field $\phi$  in the previous section with the QCD axion, 
$a$. If the amplitude of $a$ is (much) smaller than $\pi f_a$, the previous argument
holds without any change.

The strong gauge coupling becomes large and perturbative QCD breaks down at the QCD scale,
$\Lambda_{\rm QCD}$.
The QCD axion acquires a nonzero mass from non-perturbative effects of QCD, and its
mass is related to the topological susceptibility $\chi(T)$ as
\begin{equation}
\laq{massQCD}
m_a^2(T) = \frac{\chi(T)}{f_a^2}.
\end{equation}
The temperature dependence of $\chi(T)$ was estimated by several groups using lattice QCD~\cite{
Berkowitz:2015aua,Bonati:2015vqz,Petreczky:2016vrs,Borsanyi:2016ksw,Frison:2016vuc,Taniguchi:2016tjc
}, and the axion mass is parametrized as
\begin{equation}
\laq{mass}
m_a(T) \;\simeq\;
\begin{cases}
\displaystyle{\frac{\sqrt{\chi_0}}{ f_a}} \left(\frac{T_{\rm QCD}}{T}\right)^n & T \gtrsim T_{\rm QCD}\vspace{3mm}\\
\displaystyle{5.7 \times 10^{-6} \(\frac{10^{12}\GEV}{f_a}\)  {\rm eV}}& T\lesssim  T_{\rm QCD}
\end{cases},
\end{equation}
where the exponent is given by $n \simeq 4.08$~\cite{Borsanyi:2016ksw}, and
we adopt $T_{\rm QCD}\simeq 153 \MEV$ and $\chi_0 \simeq \(75.6 \MEV\)^4$.

The QCD axion starts to oscillate around the minimum when its mass becomes comparable to the Hubble parameter $H$.
The axion abundance is given by~\cite{Ballesteros:2016xej}
\begin{eqnarray}
\Omega^{}_a h^2 
\,\simeq\, 
0.35 \left(\frac{\theta_*}{0.001}\right)^{2}\times
\begin{cases}
\displaystyle 
\left(\frac{f_a}{3\times 10^{17}\,{\rm GeV}}\right)^{1.17} 
& f_a \,\lesssim\, 3 \times 10^{17}\,{\rm GeV} \vspace{3mm}\\
\displaystyle 
\left(
\frac{f_a}{3\times 10^{17}\,{\rm GeV}}\right)^{1.54}
& f_a \,\gtrsim\, 3 \times 10^{17}\,{\rm GeV}
\end{cases},
\label{QCD_axion_realignment_abundance}
\end{eqnarray}
where the initial misalignment angle is defined by $\theta_* \equiv {a_{\rm *} / f_{ a}}$.
For $\theta_* = 1$, one finds that $f_a$ should be less than about $10^{12}\GEV$ to avoid the
overabundance of the axion dark matter, $\Omega_a h^2 \lesssim 0.12$~\cite{Ade:2015xua}.
This is nothing but the upper bound of the ordinary axion window (\ref{aw}).
In deriving this upper bound, $H_{\rm inf} \gg \Lambda_{\rm QCD}$ is implicitly assumed. 

If $H_{\rm inf} \lesssim \Lambda_{\rm QCD}$, on the other hand, the axion acquires its mass during inflation.\footnote{When $H_{\rm inf} \sim \Lambda_{\rm QCD}$, the axion mass is expected to be modified from \Eq{mass} due to the difference between the de Sitter and flat space. However, this does not change the following argument significantly.}
If the inflation lasted sufficiently long (as in the case of eternal inflation),
the distribution of the axion field value follows the Bunch-Davies distribution. Thus, the initial misalignment angle  is bounded
by
\begin{equation}
|\theta_*| \lesssim \epsilon \sqrt{3 \over 8 \pi^2} {H_{\rm inf}^2 \over f_a m_a(T_{\rm inf})},
\end{equation}
where
\begin{equation}
T_{\rm inf}= {H_{\rm inf}\over 2\pi}
\end{equation}
is the Gibbons-Hawking temperature during the inflation \cite{Gibbons:1977mu},
and we have included the fine-tuning parameter, $\epsilon$.
It implies that, if $H_{\rm inf}$ is much smaller than $ \Lambda_{\rm QCD}$, $\theta_*$ becomes much smaller than $\O(1)$, 
\begin{equation} 
|\theta_*| \lesssim  0.0034\, \epsilon {\({H_{\rm inf} \over 10\MEV} \)^2} ~~~{{\rm for}~~T_{\rm inf} \lesssim T_{\rm QCD}}.
\end{equation}
Thus, the axion abundance is suppressed in the low-scale inflation, which relaxes the upper bound on the
axion window.

Note that we have assumed in the above argument that the strong CP phase during inflation is same as in the current vacuum.
This is not an unreasonable assumption because all the particles heavier than $H_{\rm inf}$ can be integrated out. In particular, the Higgs or the other heavy moduli are stabilized at their 
minima during inflation. However, if there is another light axion-like particle coupled to gluons, or if the inflaton itself is coupled to gluons, such an assumption does not necessarily hold.

We have numerically solved the equation of motion for the QCD axion field to estimate 
the axion abundance as a function of the Hubble parameter during inflation.
Our result is shown in Fig. \ref{fig:1}. The regions above the lines are excluded because the axion
is overproduced for the fine-tuning parameter, $\epsilon = 1, 0.126, 0.0125$, and $0.00125$ from left to right. 
The vertical lines correspond to the upper bounds of the ordinary axion window for $\theta_* = 1, 0.1, 0.01$, and $0.001$
from left to right. 
One can see that even $f_a \sim 10^{18}$\,GeV is allowed and the QCD axion can 
explain dark matter or a fraction of dark matter for low-scale inflation with $H_{\rm inf}< \O(1)\MEV$ without any fine-tuning of initial
angle.\footnote{Note that $6\times 10^{-13}{\rm\,eV} \lesssim m_a \lesssim 10^{-11}$\,eV, i.e., 
 $6\times 10^{17}\GEV \lesssim f_a \lesssim 10^{19}\GEV$,  is disfavored by null observations of 
 the black hole superradiance effect \cite{Arvanitaki:2014wva,
Cardoso:2018tly,
Stott:2018opm}. } This is the main result of this Letter.

  \begin{figure}[t]
  \begin{center}
   \includegraphics[width=130mm]{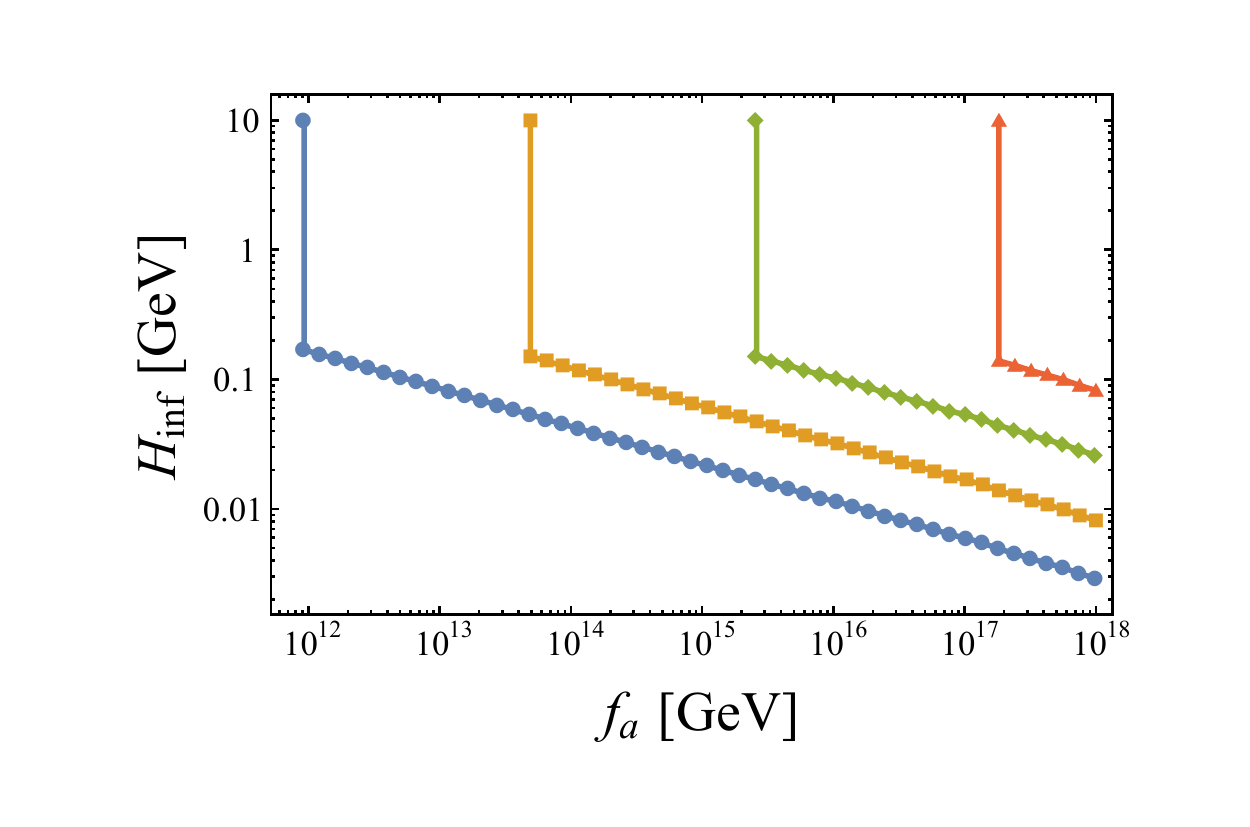}
\end{center}
\caption{The relaxed QCD axion window as a function of the inflation scale. Numerical results are shown by symbols 
 for the fine-tuning parameter, $\epsilon = 1, 0.126, 0.0125$, and $0.00125$ from left to right. 
 The vertical lines correspond to the upper bounds of the ordinary axion window for $\theta_* = 1, 0.1, 0.01$, and $0.001$
from left to right.   The region above each line
 is excluded because the axion abundance exceeds the observed dark matter abundance. 
}
\label{fig:1}
\end{figure}

Before closing this section, let us comment on the constraint of isocurvature perturbation. The quantum fluctuation of $a$ would lead to an isocurvature purturbation. The isocurvature power spectrum $\mathcal{P}_S$ is constrained to be \cite{Ade:2015lrj},
\begin{equation}
\mathcal{P}_S< \mathcal{P}_{S}^{\rm bound}\simeq 8.8\times 10^{-11},
\end{equation}
at the pivot scale $k_p=0.05{~\rm Mpc}^{-1}$. This sets an upper bound on the inflation scale as $H_{\rm inf} \lesssim {\cal O}(10^7)$\,GeV
for $f_a = 10^{12}$\,GeV. (See Ref.~\cite{Kobayashi:2013nva} for the anharmonic effect on the isocurvature perturbation.)
Since the Hubble parameter of our interest is of order the QCD scale or lower, the isocurvature bound is satisfied.

\section{Low-Scale Inflation with $H_{\rm inf}\lesssim \Lambda_{\rm QCD}$}
\lac{4}

Here we provide concrete low-scale inflation models with $H_{\rm inf}\lesssim \Lambda_{\rm QCD}$. In such low-scale inflation models,
care must be taken to achieve successful reheating.
In particular, if the reheating temperature is lower than the weak scale,
then baryogenesis becomes difficult, although not impossible. Also, if one introduces relatively strong couplings of the inflaton to the standard
model particles for successful reheating, they may spoil the flatness of the inflaton potential.

We consider a hilltop inflation model with a polynomial potential, where the inflaton $\varphi$
respects a $Z_2$ symmetry, following Ref. \cite{Nakayama:2011ri}. The potential has the following form
\begin{equation}
\laq{Vinf}
V_{\rm inf}(\varphi)=  V_0 -{m_0^2\over 2} \varphi^2 -{\kappa\over 2n} {\varphi^{2n}\over M^{2n-4}}+ {\l\over 2m} {\varphi^{2m}\over M^{2m-4}},
\end{equation}
where $M$ is a cut-off scale, and $m$ and $n$ are integers satisfying $m>n$. For simplicity, let us take $m_0 \simeq 0$ and we will return to the case of $m_0 \ne 0$ later. 
The last term stabilizes the inflaton at \begin{equation} \laq{phimin}
\varphi= \varphi_{\rm min}= \({\kappa \over \l}\)^{1\over 2(m-n)} M.
 \end{equation}
The vanishingly small cosmological constant in the present vacuum
 implies 
 \begin{align} V_0= &\({m-n\o2mn}\)\({\kappa^m \over \l^n }\)^{1\over m-n}M^4 \non \\
=&\({m-n\over 2mn}\)\kappa \varphi_{\rm min}^{2n}M^{4-2n}.
\end{align}

The inflation occurs in the vicinity of the origin where the inflaton potential is very flat. 
Inflation ends when one of the slow roll parameters, $\eta\equiv M_{\rm pl}^2V''_{\rm inf}/V_{\rm inf}$, becomes equal to 
(minus) unity,\footnote{The slow-roll parameter $\eta$ should not be confused with the conformal time in Sec. \ref{BD}.} 
where the prime denotes the derivative with respect to $\varphi$.
Solving $\eta = -1$, one obtains
\beq
\varphi=\varphi_{\rm end} \simeq \left[ {m-n\over 2mn} {1 \over 2n-1} \right]^{1\over 2(n-1)} \varphi_{\rm min}^{n\over n-1} M_{\rm pl}^{-{1\over n-1}},
\eeq
where we used the fact that in these models, $\varphi_{\rm end}
\ll M$.  The e-folding number is calculated as
\beq
N_* \simeq -\int_{\varphi_*}^{\varphi_{\rm end}}{d\varphi {3H_{\rm inf}^2 \over V'_{\rm inf}  }}\simeq {V_0  M^{2n-4} \over \kappa M_{\rm pl}^2 2(n-1)} {1\over \varphi_*^{2(n-1)}} \simeq -{2n-1 \over 2n-2} \eta_*^{-1} .
\eeq
Here and in what follows the subscript $*$ denotes the value evaluated at horizon exit of the CMB scales; in the second equality we have assumed $\varphi_*^{2(n-1)}\ll \varphi_{\rm end}^{2(n-1)}$; $N_*$ is the e-folding number during the slow-roll from the field value $ \varphi_*$ to $\varphi_{\rm end}$ (Notice that $N_*$ is smaller than $N$ in Sec.~\ref{BD}).
The spectral index is given by
\begin{equation}
\laq{nsform}
n_s \simeq1+2\eta_* \simeq 1-\({2n-1 \over n-1}\) {1\over N_*}.
\end{equation}
When $n=2$ the inflation model is reduced to the quartic hilltop inflation. In this case
the spectral index is known to be too small  to be consistent with Planck data for $N_*\lesssim 40$ \cite{Ade:2015lrj}.\footnote{
The typical value of $N_*$ is about $40$ for the low-scale inflation with $H_{\rm inf} \sim \Lambda_{\rm QCD}$.}
The spectral index can be increased by including e.g. a small $Z_2$ breaking linear term \cite{Takahashi:2013cxa}, or a 
Coleman-Weinberg potential \cite{Linde:2005ht,Nakayama:2012dw}.
As emphasized in Ref.~\cite{Takahashi:2013cxa}, such corrections to the inflaton potential slightly changes the inflaton dynamics
in such a way that the inflaton field value at the horizon exit of the CMB scale becomes closer to the origin where $|\eta_*|$
is smaller. To this end, the inflaton mass term $m_0$ is not so effective, because it not only changes the inflaton dynamics, but also 
contributes to $\eta$ in the wrong direction, and these two effects are more or less canceled. 
In any case, one can easily increase the spectral index to give a better fit to the Planck data, and 
the inclusion of these effects do not alter the following argument on the reheating.

The CMB normalization condition is given by~\cite{Peiris:2003ff,Ade:2015lrj}
\beq
A_s \equiv \Delta_{\cal R}^2 \simeq \left. \frac{V_{\rm inf}^3}{12 \pi^2 (V'_{\rm inf})^2 M_{\rm pl}^6}\right|_{\varphi=\varphi_*} \simeq 2.2 \times 10^{-9}.
\eeq
This fixes $\kappa$ as
\begin{align}
\kappa \simeq  2.6 \times 10^{-7} \(\frac{1}{2(n-1) N_*}\)^{\frac{2n-1}{n-1}}\(\frac{2mn}{m-n}\)^{\frac{n-2}{n-1}}
\(\frac{\varphi_{\rm min}}{M}\)^{-\frac{2n(n-2)}{n-1}} \(\frac{M_{\rm pl}}{M}\)^{\frac{2(n-2)}{n-1}},
\end{align}
which is reduced to 
\begin{equation}
\laq{kappa}
\kappa\simeq 5\times 10^{-13} \({40\over N_*}\)^3
\end{equation}
for $n=2$, independent of $\varphi_{\rm min}$ and $M$.

Finally the Hubble parameter during inflation and the inflaton mass at the potential minimum are given by
\begin{align}
H_{\rm inf} &\simeq  2.9 \times 10^{-4} \(\frac{m-n}{2mn}\)^{\frac{1}{2(n-1)}} 
\(\frac{1}{2(n-1) N_*}\)^{\frac{2n-1}{2(n-1)}} \left(\frac{\varphi_{\rm min}^n}{M_{\rm pl}}\)^{\frac{1}{n-1}},\\
m_\varphi^2 &=V''(\varphi_{\rm min}) = 2 (m-n) \kappa \frac{\varphi_{\rm min}^{2(n-1)}}{M^{2(n-2)}}\non \\
&\simeq 2.6 \times 10^{-7} 
\(\frac{ (m n)^{n-2} (m-n)}{  4  \(\(n-1\)N_*\)^{2n-1}}\)^{\frac{1}{n-1}}  \(\frac{\varphi_{\rm min}}{M_{\rm pl}}\)^{\frac{2}{n-1}} M^2_{\rm pl}.
\end{align} 
Notice that both $H_{\rm inf}$ and $m_\varphi$ depend only  on $\varphi_{\rm min}$.
We show $H_{\rm inf}$ and $m_\varphi$ as a function of $\varphi_{\rm min}$ in the cases of 
$(n,m) = (2,3)$ and $(2,4)$ in Fig.~\ref{fig:2} and Fig.~\ref{fig:3}, respectively.
One can see that $H_{\rm inf}= \O(1-100)\MEV$ and $m_{\varphi}= \O(10^{5}-10^{6})\GEV$ are realized 
for  $\varphi_{\rm min}= \O(10^{11}-10^{12})\GEV$.\footnote{The cut-off scale $M$ is of order $M_{\rm pl}$ for $\lambda=\O(1)$ with $m=3$. }

Characteristically for low-scale inflation models, the value of the
slow-roll parameter $\epsilon \equiv {1 \over 2} M_{\rm pl}^2
(V_{\rm inf}'/V_{\rm inf})^2$, and hence the tensor/scalar ratio $r$, are extremely
small.  In particular
\beq
 \epsilon \simeq {H_{\rm inf}^2 \over 8 \pi^2 M_{\rm pl}^2 \Delta_{\cal R}^2
    } \simeq 5.8 \times 10^6 \left( H_{\rm inf} \over
    M_{\rm pl} \right)^2 \simeq 1.0 \times 10^{-30} \left( {H_{\rm inf }
    \over \hbox{1 GeV}}\right)^2 ,
\eeq
which leads immediately to
\beq
 r \simeq 16 \epsilon \simeq 9.2 \times 10^7  \left( H_{\rm inf} \over
    M_{\rm pl} \right)^2 \simeq 1.6 \times 10^{-29} \left( {H_{\rm inf }
    \over \hbox{1 GeV}}\right)^2 .
\eeq

  \begin{figure}[!t]
  \begin{center}
   \includegraphics[width=105mm]{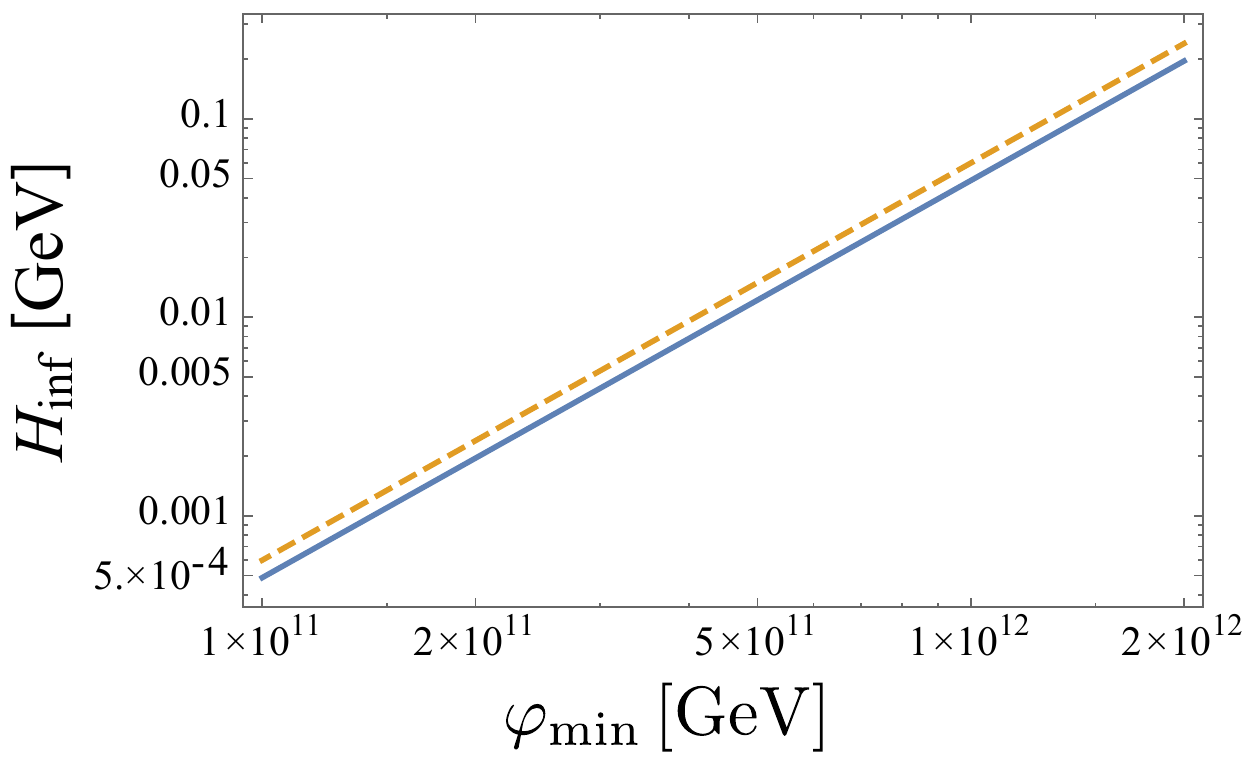}
    \end{center}
\caption{ The Hubble parameter during inflation as a function of the inflaton field value at potential minimum. 
The blue solid and orange dashed lines correspond to the cases with $(n,m) = (2,3)$ and $(2,4)$, respectively. 
The e-folding number is taken to be $N_*=40$.
}
\label{fig:2}
\end{figure}

  \begin{figure}[!t]
  \begin{center}
   \includegraphics[width=130mm]{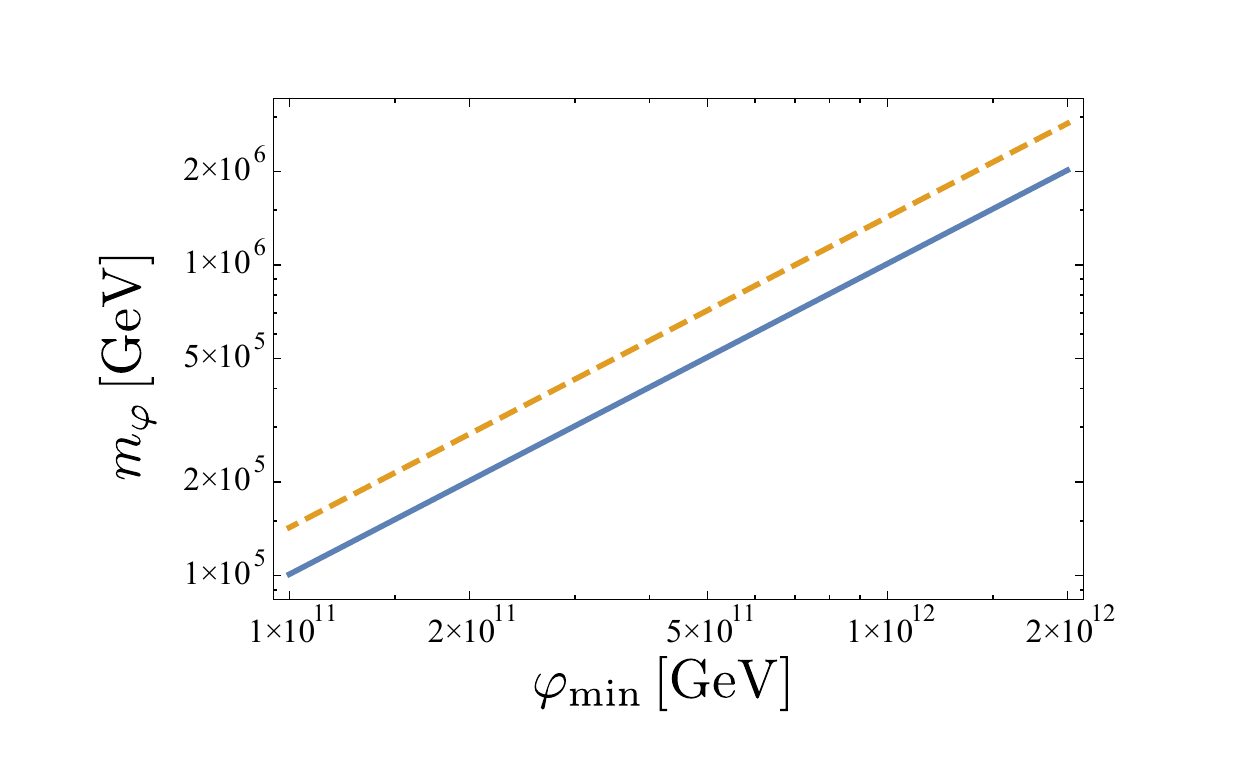}
    \end{center}
\caption{ The inflaton mass  as a function of the inflaton field value at potential minimum. 
The blue solid and orange dashed lines correspond to the cases with $(n,m) = (2,3)$ and $(2,4)$, respectively. 
The e-folding number is taken to be $N_*=40$.
}
\label{fig:3}
\end{figure}

For successful reheating we introduce  right-handed neutrinos $\nu_{R i}$ which couple to the inflaton with\footnote{
The inflaton can be identified with the B-L Higgs field. See Refs.~\cite{Nakayama:2012dw,King:2017nbl,Antusch:2018zvu} 
for detailed studies of such inflation model.
}
\begin{equation}
\sum_{i=1}^{3}{y_{N i} \over \sqrt{2}}\varphi \overline{\nu}_{Ri}^c \nu_{Ri}.
\end{equation}
Then the inflaton decay rate to a pair of right-handed neutrinos is given by 
\begin{equation}
\Gamma_{\varphi}\simeq  \sum_i^{N_{\rm R}^{\rm eff}}{ y_{Ni}^2 \over 8\pi} m_{\varphi},
\end{equation}
where the summation is taken over those neutrinos kinematically accessible by the inflaton decay.
Let us define the reheating temperature, $T_R$, as the temperature at the time $t_{\rm reheat}$ when the radiation and
the inflaton densities are equal.
The evolution of the inflaton and radiation can be described by the Boltzmann equations,
\begin{align}
\frac{d \rho_{\varphi}}{dt} &= - 3 H \rho_\varphi - \Gamma_\varphi \rho_\varphi,\\
\frac{d\rho_r}{dt} &= - 4 H \rho_r + \Gamma_\varphi \rho_\varphi,
\end{align}
with the Hubble parameter given by
\begin{align}
H &= \sqrt{\frac{\rho_r +\rho_\varphi}{3M_{\rm pl}^2}},
\end{align}
where $\rho_\varphi$ and $\rho_r$ are the energy densities of the inflaton and the thermal plasma, respectively.
Here we assume that the coherently oscillating inflaton behaves as matter, and the produced right-handed neutrinos immediately decay into the Higgs bosons and leptons.\footnote{This assumption is valid around the time of $H \sim \Gamma_\varphi$ for the neutrino Yukawa
couplings inferred by the seesaw mechanism \cite{Yanagida:1979as,GellMann:1980vs,Minkowski:1977sc}
with the light neutrino mass $\simeq 0.05\EV$.}
By solving the above equations numerically, we have found that for $\Gamma_\varphi \ll H$, the radiation energy density at $t=t_{\rm reheat}$ is $0.168 \cdot 3 M_{\rm pl}^2 \Gamma^2_\varphi$, which corresponds to a reheating temperature
\begin{align}
T_R 
&\simeq 0.64 \left({ 90  \over \pi^2 g_*} \right)^{1\over 4} \sqrt{M_{\rm pl}\Gamma_{\varphi}}\non\\
& \simeq 10 \TEV \left({106.75 \over g_*}
\right)^{1\over 4} \left({y_N\over 10^{-7}}\right) \left({m_{\varphi} \over 5 \times 10^5\GEV}\right)^{1\over 2}\left({N_R^{\rm eff}\over 2}\right)^{1/2},
\end{align}
where we have assumed $y_{Ni}=y_N$ for the kinematically allowed neutrinos.
Notice that the right-handed neutrinos are almost massless during inflation due to the $Z_2$ 
symmetry.
The right-handed neutrino acquires a mass in the present universe as
 \begin{equation} m_{\nu_R i}= \sqrt{2} y_{N i} \varphi_{\rm min}\simeq 70 \TEV \({y_{Ni} \over 10^{-7}}\)\({{\varphi}_{\rm min} 
 \over 5\times 10^{11}\GEV}\). \end{equation}
In this case, the perturbative decay is kinematically allowed and the reheating temperature 
can be higher than the electroweak scale. 
This implies that we could have successful baryogenesis via (non-thermal)
resonant leptogenesis with $N_{R}^{\rm eff}\geq 2$\cite{
Pilaftsis:1997dr,
Buchmuller:1997yu,
Pilaftsis:2003gt, Pilaftsis:2005rv,
Anisimov:2005hr,
Garny:2011hg,Dev:2017wwc} or electroweak baryogenesis.

Notice that the inflaton potential receives a radiative correction through the neutrino Yukawa interactions,  $\delta
V_{\rm inf} \simeq  {y_N^4 \over  16\pi^2 } \varphi^{4} \log \varphi$, which is negligible compared to the tree-level quartic coupling 
in the case of $n=2$.  However, one can introduce a heavier right-handed neutrino that does not contribute to the decay of the inflaton,
but its large Yukawa coupling generates a Coleman-Weinberg potential that increases the spectral index to be consistent with observation \cite{Nakayama:2012dw}.

So far, we have discussed low-scale inflation where the reheating proceeds via  perturbative decay. 
When the inflaton couples to standard-model particles strongly enough,  the inflaton may dissipate its energy 
efficiently through scattering with the ambient thermal plasma. This leads to a rather high
reheating temperature even for a relatively low inflation scale. 
In particular, Daido and two of the present authors (FT and WY) recently studied an inflation model where an axion-like particle plays 
the role of both the inflaton and dark matter \cite{Daido:2017wwb,Daido:2017tbr}. In this scenario, the reheating proceeds 
through thermal scatterings with photons and weak gauge bosons, and the inflation scale is extremely low,
$H_{\rm inf}=\O( 0.01-1)\EV$. For such an extremely low-scale inflation, the QCD axion abundance is negligible.

\section{Discussion and Conclusions}

So far, we have shown that the QCD axion window can be enlarged with low-scale inflation.
As a matter of fact, the abundance of axion-like particles can be similarly 
suppressed. In contrast to the QCD axion, their masses are usually 
assumed to be independent of time, which makes it much easier to suppress their 
abundances. In particular, it would be interesting to study cosmology with many light axion-like 
fields that follow the Bunch-Davies distribution.

 In the  low-scale  model studied above, the inflaton potential respects the $Z_2$ symmetry, and so the origin is the
symmetry-enhanced point. Therefore, under reasonable assumptions, the inflaton naturally sits at the origin
before the last inflation starts. This is the case e.g. if the inflaton acquires a Hubble-induced mass through its coupling to the
Ricci curvature during an era of power-law inflation that might precede the final inflation.

We have shown that the upper bound on the axion window can be significantly relaxed in low-scale inflation 
with the Hubble parameter smaller than the QCD scale. This is because, if the low-scale inflation lasted long enough, 
the axion initial misalignment angle follows the Bunch-Davies distribution peaked at the strong CP conserving minimum.
As a result, the axion overproduction problem is significantly relaxed. 
We have also provided a concrete low-scale inflation model with successful reheating, where
the QCD axion explains dark matter and the baryon asymmetry can be generated via resonant leptogenesis.

\vspace{5mm}

{\it Note added:}
While preparing this Letter, we found Ref.~\cite{Graham-Scherlis} which
overlaps with the present work. Compared to Ref.~\cite{Graham-Scherlis}, we focused more on building a concrete QCD-scale inflation model
with successful reheating and pointed out that the absence of any other light scalars (including the inflaton) contributing 
to the strong CP phase is crucial for the present mechanism to work.

\paragraph*{Acknowledgments:} 
F.T. thanks the hospitality of MIT Center for Theoretical Physics where this work was done. 
This work is supported by JSPS KAKENHI Grant Numbers JP15H05889 (F.T.), 
JP15K21733 (F.T.), JP17H02878 (F.T.), and JP17H02875 (F.T.), Leading Young Researcher Overseas
Visit Program at Tohoku University (F.T.),
and by World Premier International Research Center Initiative (WPI Initiative), 
MEXT, Japan (F.T.). A.H.G. is supported in part by the U.S. Department of Energy under grant 
Contract No. DE-SC0012567.

\end{document}